\documentclass[usegraphicx]{mn2e}
\usepackage{subfigure}
\usepackage{amsmath}
\usepackage{amssymb}
\usepackage{epstopdf}
\usepackage{graphics,graphicx}
\usepackage{epsfig}

\newcommand {\apgt} {\ {\raise-.5ex\hbox{$\buildrel>\over\sim$}}\ }
\newcommand {\aplt} {\ {\raise-.5ex\hbox{$\buildrel<\over\sim$}}\ } 
\newcommand {\degree}{$^{\circ}$}

\title[The low spin of the first extragalactic microquasar]
{The low or retrograde  spin of the first extragalactic microquasar:
implications for Blandford-Znajek powering of jets}

\author[M.Middleton et al.]
{Matthew J. Middleton$^{1}$, James C.~A. Miller-Jones$^{2}$ and Rob P. Fender$^{3, 4}$\\
1. Astronomical Institute Anton Pannekoek, Science Park 904,1098 XH, Amsterdam, Netherlands\\
2. International Centre for Radio Astronomy Research - Curtin University, GPO Box U1987, Perth, WA 6845, Australia\\
3. Physics and Astronomy, University of Southampton, Highfield, Southampton SO17 1BJ, UK\\
4. Department of Physics, Oxford University, Denys Wilkinson Building, Keble Road, Oxford OX1 3RH, UK
}

\pagerange{\pageref{firstpage}--\pageref{lastpage}} \pubyear{2011}
\long\def\symbolfootnote[#1]#2{\begingroup\def\thefootnote{\fnsymbol{footnote}}\footnote[#1]{#2}\endgroup} 
\def\ga{\mathrel{\hbox{\rlap{\hbox{\lower4pt\hbox{$\sim$}}}{\raise2pt\hbox{$>$}}
}}}

\begin{document}

\topmargin = -0.5cm

\maketitle

\label{firstpage}

\begin{abstract}

Transitions to high mass accretion rates in black hole X-ray binaries
are associated with the ejection of powerful, relativistically-moving
jets. The mechanism that powers such events is thought to be linked to
tapping of the angular momentum (spin) of the black hole, the 
rate of accretion through the disc or some combination of the
two. We can attempt to discriminate between these different possibilities by comparing proxies for jet power with spin estimates. Due
to the small number of sources that reach Eddington mass accretion
rates and have therefore been suggested to act as `standard candles', there has been much recent
debate as to whether a significant correlation exists between jet power and black hole spin. We perform
continuum fitting to the high-quality, disc-dominated {\it XMM-Newton}
spectra of the extragalactic microquasar discovered in M31. Assuming prograde spin, we find that, for sensible constraints the spin is always very low ($a_*$ $\le$
0.15 at 3$\sigma$). When combined with a proxy for jet power derived from the
maximum 5~GHz radio luminosity during a bright flaring event, we find
that the source sits well above the previously reported, rising
correlation that would indicate that spin tapping is the dominant
mechanism for powering the jets, i.e. it is too `radio loud' for such
a low spin. The notable exceptions require the inclination to be improbably small or the jet to be very fast. We investigate whether this could be a by-product of selecting prograde-only spin, finding that the data statistically favour a substantially retrograde spin for the same constraints ($a_*$ $\le -0.17$ at 3$\sigma$). Although theoretically improbable, this remarkable finding could be confirmation that retrograde spin can power such jets via spin-tapping, as has been suggested for certain radio quasars. In either case this work demonstrates the value of studying local extragalactic microquasars as a means to better understand the physics of jet launching.

\end{abstract}

\section{Introduction}

Extensive X-ray observations of the few dozen Galactic transient black
hole X-ray binaries (BHBs, see the review of Remillard \& McClintock
2006) have shown similar patterns of spectral evolution over the
course of an outburst, with sources tracing a characteristic shape in
hardness and intensity (Homan et al. 2001; Fender, Gallo \& Belloni
2004). The latter is generally parameterised as a ratio of bolometric
luminosity to the Eddington luminosity, $L_{\rm Edd}$. At low Eddington fractions (below $\sim0.5\% L_{\rm Edd}$; Dunn et al. 2010), and at higher Eddington fractions when in the hard X-ray spectral state, the
X-ray spectrum is dominated by a hard component of emission potentially
arising from Compton up-scattering of photons from a weak accretion
disc in an optically thin plasma of hot, thermal electrons (see Done,
Gierlinski \& Kubota 2007 for a review). An alternative suggestion is
that this component is the high-energy extension of the non-thermal
synchrotron emission seen from the lowest observable frequencies up to
at least the IR band, where it breaks to become optically thin
(e.g. Markoff, Falcke \& Fender 2001; Gandhi et al. 2011). Although there is good evidence for the extension of the synchrotron emission to X-ray energies at low mass accretion rates ($\sim0.1\% L_{\rm Edd}$; see Russell \& Shahbaz 2014), it is unlikely to make a dominant contribution to the X-ray bandpass at higher accretion rates (e.g. Malzac, Belmont \& Fabian 2009). X-ray emission aside, the low (radio-IR) frequency emission is well-accepted to arise from synchrotron radiation from highly relativistic electrons moving at what are assumed to be relatively low bulk Lorentz factors ($\Gamma$, Casella et al. 2010) in a steady, compact,
collimated outflow, or jet; BHBs showing such jets (thought to be ubiquitous) are often termed `microquasars'. The details of the jet launching and
particle acceleration are still uncertain, however, it is clear that
the properties of the jet and inflow are causally connected, with the radio and X-ray emission being well correlated
(e.g. Corbel et al. 2003, 2013; Falcke, K{\"o}rding, \& Markoff 2004; Merloni, Heinz \& Di
Matteo 2003; Plotkin et al. 2012). At high mass accretion rates, the X-ray spectral state changes, with the emission becoming dominated by thermal disc emission.  During this spectral state transition, the properties of the outflow change dramatically, with discrete, ballistically-moving ejecta launched at high bulk Lorentz factors (assumed to be $\Gamma\sim$2--5; Fender et al.\ 2004, although see Fender 2003 for a discussion on constraining $\Gamma$).  The compact jets become quenched, and are not re-established until the reverse transition (at lower luminosities due to the spectral hysteresis), when the hard power-law X-ray emission is re-established.

The `ballistic jets' seen at the peak of the outburst are typically much brighter at GHz frequencies than the `persistent jets' associated with hard X-ray emission (see Fender et al. 2009 and references therein for details), and are the likely analogues of discrete ejecta seen from
radio loud quasars, also thought to accrete at very high
rates. %As a result, BHBs in this phase are referred to as microquasars (Mirabel et al. 1992).

\begin{figure*}
\includegraphics[width=3in]{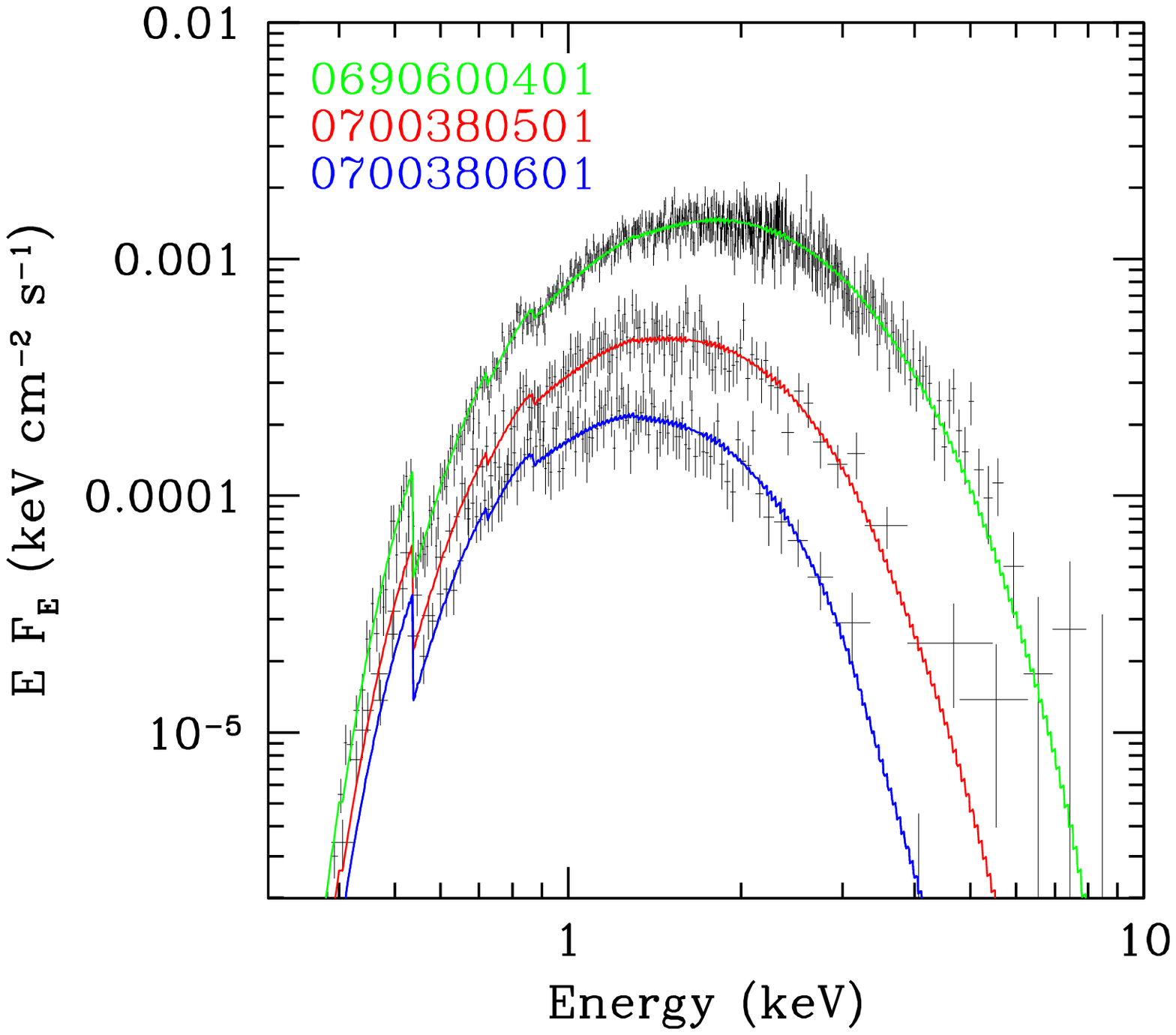}
    \hspace{1cm}
\includegraphics[width=2.65in]{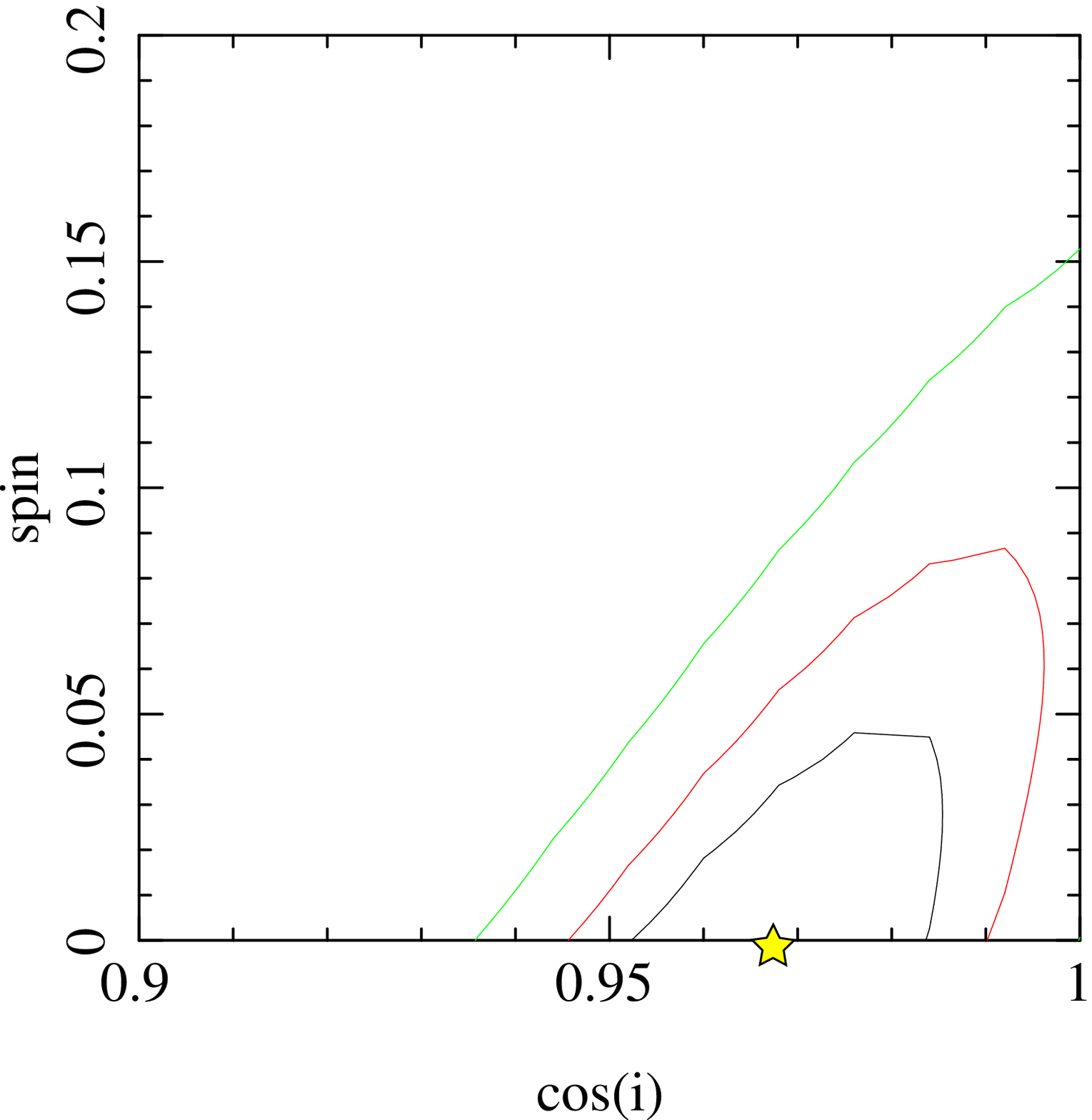}
%\vspace{-0.5cm}
\caption{{\it Left: XMM-Newton} EPIC PN spectra from three observations
  taken during the decay phase (Barnard et al. 2013, M+13). We fit these simultaneously using a model in {\sc xspec}
  comprising neutral absorption ({\sc tbabs} with column density
  tied between datasets) and disc emission that incorporates vertical
  and radiative transport together with relativistic smearing ({\sc
    bhspec}). We find that for an inferred mass of 10$M_{\odot}$ we
  obtain $a_*< 0.06$ (90\%) with the spin and inclination free. {\it Right:} As we have allowed inclination to be free in our fitting we
  also show the 1, 2 and 3$\sigma$ contours (black, red and green respectively) for this parameter space as well (the star corresponds to the best-fitting value) .}
\end{figure*}

As powerful jet events appear during state transitions at
high mass accretion rates, understanding their properties and how they
are launched has cosmological relevance, since accretion rates
approaching (or exceeding) Eddington are invoked for the growth
of supermassive black holes in quasars (QSOs; Fan et
al. 2003). Whilst there is still uncertainty as to whether this growth
was steady or chaotic (see Fanidakis et al. 2010 for a discussion on
the modes), the latter growth path would give rise to many epochs of
outburst and associated ballistic jet ejections. Besides the obvious
implications such accretion rates have for providing ionising flux at
the epoch of re-ionisation, the outflows associated with each outburst
will have removed matter, energy and angular momentum from the
accretion flow, with the expelled, hot material interacting with the
developing local environment (see Fabian 2012 for a recent review of AGN feedback).

\begin{table*}
\begin{center}
\begin{minipage}{145mm}
\caption{{\sc xspec} fitting results for prograde spin.}
\begin{tabular}{c|c|c|c|c|c|c}
  \hline

% after \\: \hline or \cline{col1-col2} \cline{col3-col5} ...
 $N_{\rm H}$ ($\times$10$^{22}$cm$^{-2}$) & L/L$_{\rm Edd}$ Obs1  & L/L$_{\rm Edd}$ Obs2 & L/L$_{\rm Edd}$ Obs3 &  i
(\degree) & a$_*$ & $\chi^2$/dof\\
\hline
 0.41 $\pm$ 0.01 & 0.199 $\pm$ 0.002 &  0.077 $\pm$ 0.002  &  0.041 $\pm$ 0.001 & 32 & 0 ($<$ 0.004) & 910/764\\

 0.39 $\pm$ 0.01 &  0.209 $\pm$ 0.002 & 0.079 $\pm$ 0.002  &  0.042 $\pm$ 0.001 &  39 & 0 ($<$ 0.002) & 1109/764\\ 

\hline
0.45 $\pm$ 0.01  & 0.187$^{+0.002}_{-0.003}$ & 0.074 $\pm$ 0.002 & 0.041 $\pm$ 0.001 & 14.6 $^{+3.5}_{-5.7}$ & 0  ($<$ 0.056) & 781/763\\
\hline 

\end{tabular}
Notes: {\it Top:} Best fitting spectral parameters (and 90\% errors) across the three
tied {\it XMM-Newton} datasets (denoted by Obs 1-3) using the {\sc xspec} model {\sc tbabs*bhspec}. In each case the inclination has been
determined by comparing the brightness temperature of the minute-timescale radio variability with the analogous oscillation events seen in GRS~1915+105, assuming a boosting factor of $\sim$15 and a mass of
10~$M_{\odot}$ for the M31 source. The normalisation is fixed at 1.68$\times$10$^{-4}$ as
the distance to M31 is well constrained. {\it Bottom:} Best-fitting
model parameters with both inclination and spin as free parameters.

\end{minipage} 

\end{center}
\end{table*}

A strict determination of how these ejecta are launched, and the
relative amounts of matter and energy extracted from the accretion
flow, requires simultaneous observations of both inflow and
outflow. The timescales for such studies are impractical for even the
low-redshift analogues of QSOs, the narrow line Seyfert 1 galaxies,
for which the viscous timescales in the disc are of order decades to
millennia. However, the disc-jet coupling has been well studied in
Galactic BHBs, although observations are often hampered by the
distorting effects of soft X-ray absorption.  The recently confirmed
class of extragalactic microquasars (Middleton et al. 2013, M+13
hereafter) provide an avenue by which the disc and jet can be studied
simultaneously, with the prospect of lower absorption allowing a
clearer view of the accretion disc (Middleton et al. 2012; M+13) as well as a small
fractional uncertainty in the distance. However, investigations into
such sources are in their infancy and, as a result of these complicating
issues, the physics behind the launching of these jets remains to be fully understood.  

An indirect test of the launching mechanism can be
performed by measuring and comparing  proxies for the jet power and the angular momentum (usually
parameterised by the dimensionless spin) of the black hole (Fender et al. 2010; Narayan \& McClintock 2012, NM12
hereafter; Steiner, McClintock \& Narayan 2013, S13 hereafter;
Russell, Gallo \& Fender 2013). These two quantities are linked in the
Blandford-Znajek (BZ) model for jets (Blandford \& Znajek 1977), in
which the jet power is proportional to the square of the spin (in the
low spin limit). Although it is still debated as to how to measure either quantity
unambiguously, the jet power, $P_{\rm jet}$, has previously been
estimated for BHBs using e.g. the synchrotron minimum energy (Russell et
al. 2013), the normalisation of the radio/X-ray correlation (Fender et al. 2010) or the peak luminosity in a particular radio
band (NM12; S13). The latter proxy assumes a broadband synchrotron
spectral shape, predictable evolution of jet brightness, impulsive
injection of jet power and a linear relationship between radiative and
kinetic energies (see Appendix B of S13 for a derivation of this and
associated caveats), without which the scaling would be flawed.

Measuring the spin (a$_{*}$) has also been contentious. It is usually determined (although see 
 Abramowicz \& Kluzniak 2001; Kluzniak \& Abramowicz 2001 for a different approach as applied by Motta et al. 2013) either by
modelling the reflection spectrum around the Fe K${\alpha}$ line,
taking into account relativistic broadening (Fabian et al. 1989), or
else by modelling the continuum in cases when the X-ray emission is
dominated by the accretion disc (Ebisawa, Mitsuda \& Hanawa
1991). Both techniques rely on emission originating from the innermost
stable circular orbit (ISCO), which moves from $6R_{\rm g}$ for a
non-spinning, Schwarzschild black hole (a$_{*}$ = 0), down to $1.24R_{\rm g}$ when
the spin is maximal prograde (a$_{*}$ = 0.998) and out to 9$R_{\rm g}$ for maximal retrograde spin (a$_{*}$ = -0.998). Although retrograde spin can result from chaotic accretion in the growth of SMBHs (Fanidakis et al. 2011), creating such an alignment of the accretion flow is much harder in BHBs requiring either wind fed accretion or a secondary capture scenario in a dense stellar environment. Although the existence of retrograde-spins is therefore theoretically challenging and has never been unambiguously confirmed (though suspected in some FRII QSOs such as 3C120: Kataoka et al. 2007, and the Galactic BHBs, GRS~1124-68: Zhang, Cui \& Chen 1997 and Swift J1910.2-0546: Reis et al. 2013), it has been speculated that such spins could lead to extremely powerful jet events by sweeping the magnetic flux in the plunging region onto the black hole.  As the `gap' between the ISCO and BH is larger for retrograde spin, the magnetic flux trapped on the black hole can therefore be enhanced (Garofalo 2009; Garofalo, Evans \& Sambruna 2010). However,  simulations incorporating the effect of magnetic field saturation (Tchekhovskoy \& McKinney 2012) dispute this `gap model'. Instead it is proposed that where a large magnetic flux density is present in the disc, the flow is arrested (i.e. a Magnetically Arrested Disc: MAD, e.g. Bisnovatyi-Kogan \& Ruzmaikin 1974; Igumenshchev, Narayan \& Abramowicz 2003; Narayan, Igumenshchev \& Abramowicz 2004; Tchekhovskoy, Narayan \& McKinney 2011), with a common feature of such systems being that the BH magnetic flux builds to a saturation point. As a result, the plunging region is predicted to fill with magnetic flux where the inflow and outflow rates are matched, and therefore no gap is produced in the simulations. This model predicts that the BZ effect produces jets with a moderately {\it lower} efficiency at retrograde spins than at prograde spins (Tchekhovskoy \& McKinney 2012).

Initial studies combining estimated $P_{jet}$ and spin values showed no evidence for spin-powering of jets (Fender et al. 2010).  More recently, NM12 claimed evidence for a correlation between $P_{jet}$ and spin in transient jets, implying they were powered by the BZ effect.  However, this work remains controversial (see, e.g. Russell et al. 2013), as it relies on taking {\it only} those sources which approach
their Eddington limit and act as `standard candles' (see Appendix A of
S13).

Since there are so few microquasars known in our Galaxy, the sample
size is small, leaving the issue of jet powering open to debate. To
improve the statistics we include a new source by modelling the
disc-dominated spectra of the microquasar recently discovered in M31,
whose 2012 outburst reached the Eddington luminosity and was well
studied in both the radio and X-ray bands (M+13).

%\begin{figure*}
%\centering
%\mbox{\subfigure{\includegraphics[width=3in]{spectra_retro.ps}\quad
%\subfigure{\includegraphics[width=3in,angle=-90]{cont_retro.eps} }}}
%\caption{As for Figure 2 fitting with the expanded {\sc bhspec} model allowing us to test for retrograde spin. When both spin and inclination are allowed to be free we find the spin to be constrained (at 90\%) to $<$ -0.57. As before the spectral fits are shown in the left-hand plot and contours for inclination and spin parameter space on the right.}
%\end{figure*}

%\begin{figure}
%\centering
 %\includegraphics[width=\columnwidth]{spectra_retro.ps}
%\epsfig{file=spectra.ps}
%\caption{As for Figure 2 fitting with the expanded {\sc bhspec} model allowing us to test for retrograde spin. When both spin and inclination are allowed to be free we find the spin to be constrained (at 90\%) to $<$ -0.57. As before the spectral fits are shown in the left-hand plot and contours for inclination and spin parameter space on the right.} \centering
%\label{fig:l}
%\end{figure}

%SHOW DEL_CHI

\begin{figure*}
\includegraphics[width=3in]{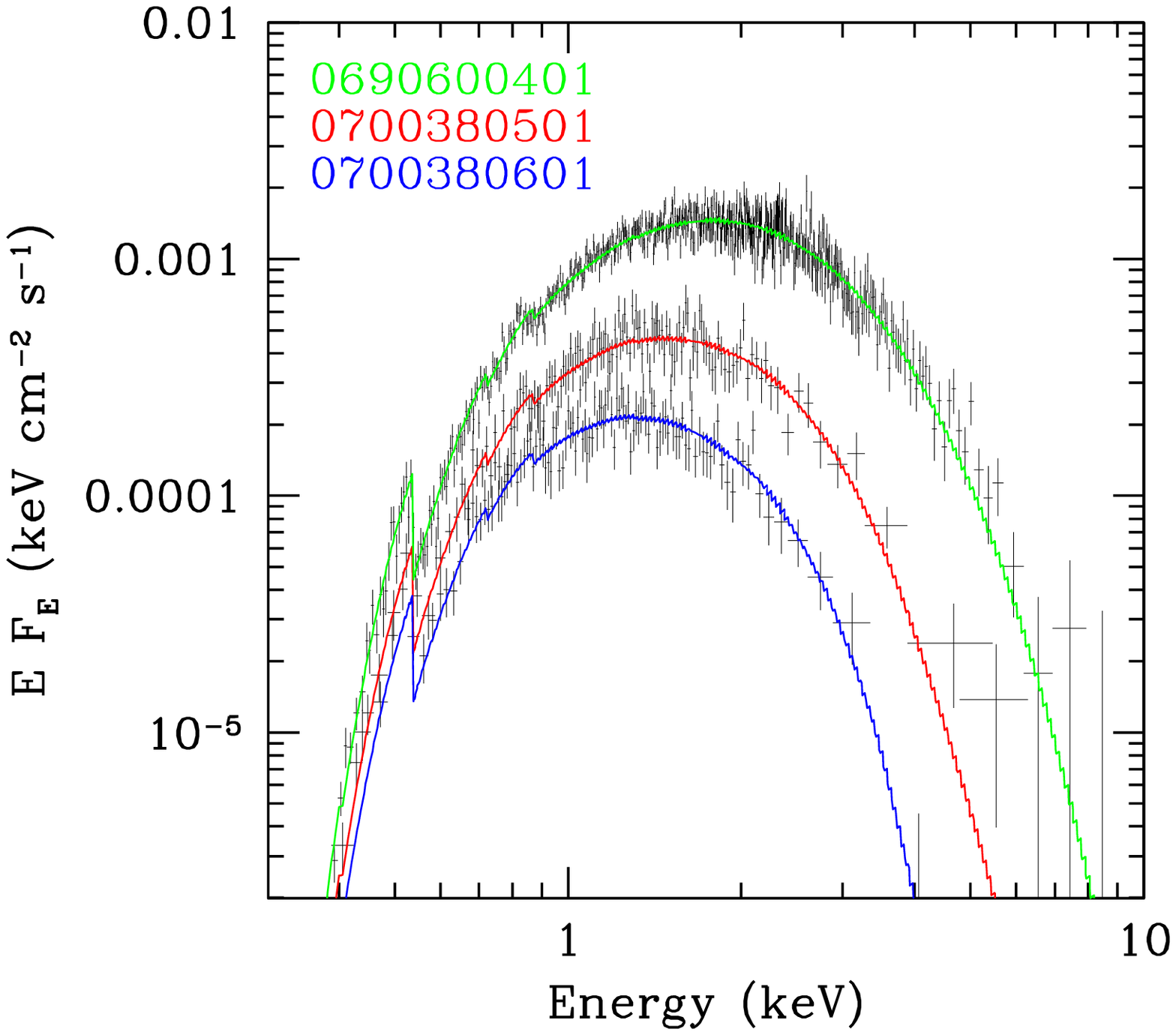}
    \hspace{1cm}
\includegraphics[width=2.65in]{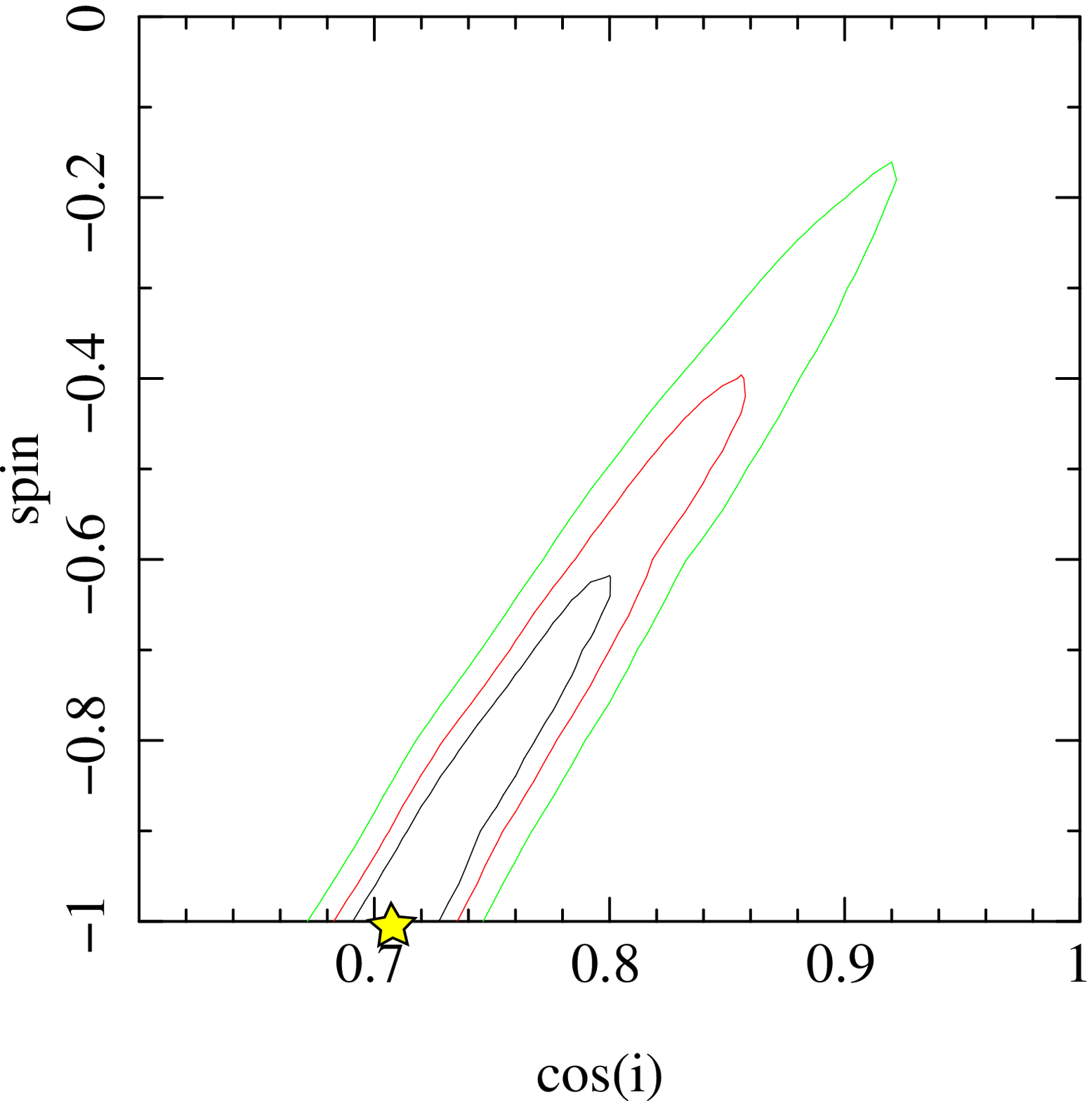}
%\vspace{-0.5cm}
\caption{As for Figure 1, showing the best fits with the expanded {\sc bhspec} model, allowing a test for retrograde spin. When both spin and inclination are allowed to be free, we find the spin to be constrained (at 90\%) to $<$ -0.57. As before, the spectral fits are shown in the left-hand plot and contours (with 1, 2 and 3$\sigma$ levels shown in black, red and green respectively) for inclination and spin parameter space on the right (again the star corresponds to the best-fitting value).}
\end{figure*}

\begin{table*}
\begin{center}
\begin{minipage}{145mm}
\caption{{\sc xspec} fitting results for retrograde spin.}
\begin{tabular}{c|c|c|c|c|c|c}
  \hline

% after \\: \hline or \cline{col1-col2} \cline{col3-col5} ...
 $N_{\rm H}$ ($\times$10$^{22}$cm$^{-2}$) & L/L$_{\rm Edd}$ Obs1  & L/L$_{\rm Edd}$ Obs2 & L/L$_{\rm Edd}$ Obs3 &  i
(\degree) & a$_*$ & $\chi^2$/dof\\
\hline
 
 0.46 $\pm$ 0.01 & 0.214 $\pm$ 0.003 & 0.085 $\pm$ 0.002  & 0.047 $\pm$ 0.001 &  32 & -0.44 $\pm$ 0.07 & 770/764\\ 
 
 0.46 $\pm$ 0.01  & 0.237 $\pm$ 0.004 & 0.094 $\pm$ 0.003    & 0.052 $\pm$ 0.002  & 39  & -0.72 $\pm$ 0.08 & 767/764\\ 
 
\hline
0.46 $\pm$ 0.01 & 0.264$^{+0.005}_{-0.039}$ & 0.105$^{+0.003}_{-0.015}$  & 0.058 $\pm$ 0.001 & 44.8 $^{+0.9}_{-9.0}$ &  -0.998  ($<$ -0.566) & 766/763\\ 
\hline 
\end{tabular}
Notes: {\it Top:} As for Table 1 using the extended {\sc bhspec} model allowing retrograde spins to be tested.
\end{minipage} 
\end{center}
\end{table*}

\section{The 2012 Outburst} 

The M31 source, XMMU J004243.6+412519, was discovered in 2012 during the rise phase of an X-ray outburst, reaching a peak (0.3-10~keV) luminosity of
$\sim$1.3$\times$10$^{39}$erg~s$^{-1}$. It was subsequently monitored
in the X-rays by {\it Swift} and in the radio by the Karl
G. Jansky Very Large Array (VLA), the Very Long Baseline Array (VLBA)
and the Arcminute MicroKelvin Imager - Large Array (AMI-LA). The
latter revealed a sequence of radio flaring events, persisting for a
few months after the peak of the outburst. This is very
similar to the behaviour seen in the 1993 December--1994 April
outburst of GRS 1915+105 (Harmon et al.\ 1997); multiple radio flares
over the course of several months during a single X-ray outburst.

\subsection{Black hole mass}

When brightest (at a 0.3-10~keV luminosity of
$\sim1.3\times10^{39}$\,erg\,s$^{-1}$), the X-ray spectrum of the
microquasar in M31 appeared to require a two-component model with a
`cold', optically-thick electron plasma reprocessing photons from a
quasi-thermal accretion disc, as seen in XRBs accreting close to the
Eddington rate (e.g. Ueda et al. 2009). At this time, the peak disc
temperature deviated strongly from the $L\propto T^{4}$ relation
expected for a thin disc (Shakura \& Sunyaev 1973), likely indicating
a change in structure (e.g. Poutanen et
al. 2007) that would only be expected close to Eddington rates. Therefore
we are confident that the source was accreting at or very close to
Eddington rates and the associated radio emission was associated with
ballistic jet ejection (rather than a persistent jet, which should be
much less luminous). Based on this X-ray
and radio phenomenology, M+13 identified the source as a microquasar
accreting at the Eddington rate with a black hole
mass close to $10M_{\odot}$. 

Following the peak of the outburst, the X-ray emission appeared disc
dominated, decaying over the course of $\sim$200 days and demonstrating the typical L$\propto$T$^{\sim4}$ behaviour seen in
outbursting Galactic BHBs , and implying a constant emitting area (e.g. Done et al. 2007). Over this
period, three pointed {\it XMM-Newton} observations were taken
(OBSIDs: 0690600401 taken on 2012-06-26, 0700380501 taken on 2012-07-28 and 0700380601 taken on 2012-08-08), yielding high-quality, disc-dominated
X-ray spectra.  

By comparing the luminosity of the source in its dimmest observed soft state (7$\times$10$^{37}$ erg s$^{-1}$) to the Eddington fraction at which Galactic BHBs are seen to transit from the soft to the hard state (Dunn et al. 2010), an estimate for the upper limit on the BH mass can be obtained. This limit corresponds to the end of our X-ray monitoring; the source may well have continued in a disc-dominated state down to lower luminosities. Whilst observed transitions have been reported as low as 0.5\% of Eddington (implying a mass limit close to 100$M_{\odot}$), the combination of poor population statistics, uncertainty in mass and distance determination for Galactic BHBs, and the different bandpasses used for the analysis, led M+13 to use the most representative value for the transition of $\approx$3\% (Dunn et al. 2010). M+13 consequently placed an upper limit on the
black hole mass of 17$M_{\odot}$. We
note that should we take this upper limit to be representative of the
mass then it becomes hard to explain the unusual source
phenomenology (both radio and X-ray), which seems to match only GRS 1915+105 when at
Eddington (we note the recently discovered BHB: IGR17091-3624 may also show such behaviour, although the distance and mass are unknown at present; Altamirano et al. 2011). In addition, we might expect that M31, having a similar accreting source population to our own Galaxy should have a similar black hole mass distribution ({\"O}zel et al. 2010), which would suggest that a mass well above 10$M_{\odot}$ is relatively improbable (though observationally not excluded). We are therefore satisfied that the lower value of
10$M_{\odot}$ is more plausible and use this throughout.
 
 \subsection{Inclination angle}
 
Although we cannot determine the inclination to the source directly, the lack of strong absorption in the X-ray spectrum implies a line of sight that does not intercept the large columns of equatorial wind expected in such a state (Ponti et al. 2012), i.e. $<$60\degree\ (based on the inclination and winds of GRO~J1655-40, e.g. Neilsen \& Homan 2012). We can obtain a more constraining, although indirect, estimate by comparing the variability brightness temperature of the
short-timescale (10s of minutes) variability in the M31 microquasar to similar flaring
events (the half-hour `oscillations') in GRS~1915+105 (Pooley \&
Fender 1997). In doing so M+13 found a factor $\approx$15 amplification in the M31 source (explicitly assuming that spin does not affect the brightness of such events). The variability brightness temperature scales as $\delta^3$, where $\delta = [\Gamma(1-\beta\cos i)]^{-1}$ is the Doppler factor.  Given a fixed jet speed and the inclination angle of GRS~1915+105, the ratio of the variability brightness temperature of the M31 source to that of the oscillation events in GRS~1915+105 (i.e. $[\delta$(M31)/$\delta(1915)]^3$) can be used to solve for the inclination. Taking the inclination of GRS~1915+105 to be 66\degree\ (Fender et al. 1999),  this corresponds to
an inclination of the M31 source of 32\degree\ for $\Gamma$ = 2 (Fender et al. 2004; NM12) and 39\degree\ for $\Gamma$ = 5 (Fender et al.\ 2004;
S13). If instead we compare to the maximum measured brightness
temperatures in other BHBs, the beaming factor could be as high as 70
(M+13), and the inclination would be considerably smaller. However, given the rather
unique phenomenology shared by GRS~1915+105 and the M31 source, it seems most appropriate
to use the former boosting factor. 

\section{Continuum fitting} 

As the distance to M31 is extremely well constrained
($d=772\pm44$~kpc; Ribas et al. 2005) and the black hole mass can be
inferred, we were able to apply the best disc models to the
{\it XMM-Newton} spectra of the decay phase and place constraints on
the black hole spin (e.g. Steiner et al. 2012). We note that the X-ray luminosities from
the three observations correspond to Eddington ratios $<30$\%, so
they should be unaffected by uncertainties in the disc behaviour that affect observations at
high Eddington ratios (McClintock et al. 2006). In addition, the low
column density along the line of sight to the source has allowed the
spectra to be modelled more accurately than for many Galactic sources,
where high columns of intervening material in the plane of the Galaxy
can obscure emission below 2~keV (e.g. Zimmermann et al. 2001).

\begin{figure*}
\includegraphics[width=6in]{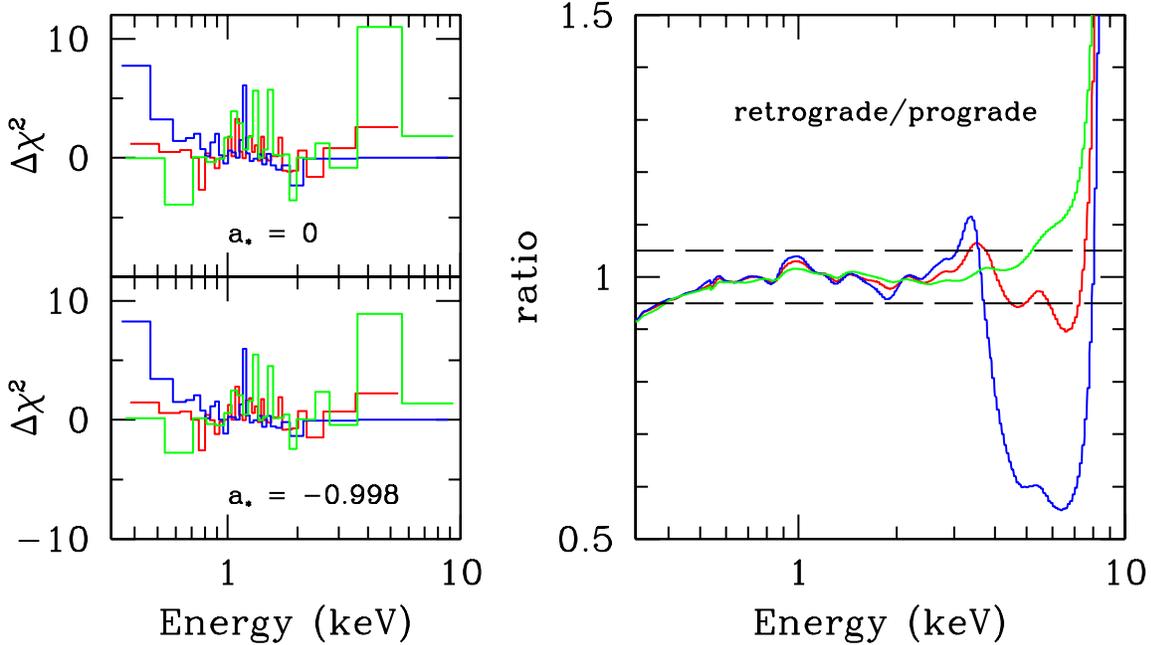}
\caption{{\it Left:} $\Delta\chi^{2}$ plots corresponding to the spectra in Figures 1 and 2 (green: 0690600401, red: 0700380501 and blue: 0700380601), re-binned for inspection purposes. {\it Right:} ratio of the best-fitting retrograde and prograde models with inclination and spin as free parameters. Although there are major discrepancies in the models at high energies, the data quality above 3~keV in the two dimmer observations is poor and so the effect on $\chi^2$ is only weak. Instead it is clear that the major changes in $\Delta\chi^{2}$ correspond to the energy range where changes in the best-fitting models are $\le$5\% (horizontal dashed lines), of the order of the model systematics. Therefore, whilst the retrograde model is statistically favoured, we cannot yet robustly confirm its validity; although if the spin is not retrograde, the large jet power would then imply that the BZ mechanism is not operating as expected in this source.}
\end{figure*}

For completeness, we re-extracted the EPIC PN spectra (the MOS data
add little in terms of statistics) using {\sc sas v12} to reproduce
raw event files. We filtered the high-energy
(10-15~keV) light curves for periods of soft proton flares and used
standard flags (=0) and patterns ($\le$4) to produce spectra from
source and background regions of 35$^{\prime\prime}$ radius (avoiding
read-out node directions for the background). Spectra were extracted
using {\sc xselect} and grouped to 25 counts/bin for chi-squared
fitting. We find that a basic disc model ({\sc diskbb}) convolved with neutral absorption by the Galactic and host ISM 
({\sc tbabs}, with abundances set by Wilms, Allen \& McCray 2000 and  the lower limit set to the Galactic column along the
line-of-sight to M31: 7$\times$10$^{20}$ cm$^{-2}$; Dickey \& Lockman
1990) can fully describe the data and we do not find a statistical requirement for any higher-energy component as seen when the source was brightest (M+13). As a result we proceeded to use the disc
model for prograde spin, {\sc bhspec} (Davis et al. 2005), to simultaneously model the three spectra and thereby obtain tighter constraints. This model includes full radiative
transfer, vertical transport and relativistic
broadening and takes black hole mass, Eddington ratio, spin, inclination and normalisation as input model parameters. The normalisation is dependent on the distance (norm = (10/d$_{\rm
  kpc})^{2}$) and so is well constrained for this source and the mass is well reasoned to be $\approx$10M$_{\odot}$ (see $\S$2.1). By taking the inclinations discussed in $\S$2.2 we therefore have the requisite inputs for {\sc bhspec}, leaving only Eddington ratio and spin as unknowns in the model fitting.

\begin{table}
\begin{center}
\begin{minipage}{55mm}
\caption{Jet power proxy values}
\begin{tabular}{c|c|c}
  \hline\hline

% after \\: \hline or \cline{col1-col2} \cline{col3-col5} ...

\multicolumn{3}{c}{Galactic BHBs}\\
\hline
Source &  $\Gamma$=2, P$_{jet}$ &  $\Gamma$=5,  P$_{jet}$\\
\hline
A0620-00 & 0.13 & 1.6\\
XTE J1550-564 & 11 & 180\\
GRO J1655-40 & 70 & 1600\\
GRS 1915+105 & 95 & 1700\\
H1743-322 & 7.0 & 140\\
\hline\hline
\multicolumn{3}{c}{M31 microquasar}\\
\hline
 $\Gamma$ & i\degree  & P$_{jet}$\\
\hline
\multicolumn{3}{c}{Fixed inclination}\\
\hline
% 2 & 0 & 32 & 41.3\\
 2 & 32 & 52\\
 %5 & 0 & 39 & 482.7\\
 5 & 39 & 810\\
 \hline
\multicolumn{3}{c}{3$\sigma$ upper limits (prograde spin)}\\ 
\hline
%2 & 0 & 0 & 5.1\\
2 & 0 & 5.0\\
%5 & 0 & 0 & 0.29\\
5 & 0 & 0.18\\
\hline
\multicolumn{3}{c}{Best-fitting (retrograde spin)}\\ 
\hline
%2 & 0 & 45 & 128\\
2 & 45 & 180\\
%5 & 0 & 45 &  1011\\
5 & 45 & 1900\\
\hline\hline 
\end{tabular}
\vspace{0.5cm}
\end{minipage} 
\begin{minipage}{80mm}
Notes: {{\it Upper:} $P_{jet}$ values (with units of GHz Jy kpc$^2$ M$_{\odot}^{-1}$ and given to 2 s.f.) for the 5 BHBs of S13, derived using the inclinations, masses, flux densities, $\alpha$ values and distances given in NM12 and S13. {\it Lower:} Bulk Lorentz factors ($\Gamma$) and inclinations (i) used to derive the jet power proxies (P$_{jet}$ to 2 s.f.) for the M31 microquasar, following the method of NM12. {\it Top:} the inclination is derived from scaling of the variability brightness temperature to the analogous events in GRS~1915+105 (i.e. "Fixed inclination"). {\it Middle:} The inclination of 0\degree\ is the 3$\sigma$ upper limit to the fit using prograde spin. {\it Bottom:} The inclination of 45\degree\ is the best-fitting value where the spin is allowed to be retrograde.} 
\end{minipage} 
\end{center}
\end{table}

The absorption column was tied between observations, {\sc
  bhspec} normalisations fixed to $1.68\times10^{-4}$, mass fixed at $10M_{\odot}$ and inclination fixed
initially to be 32\degree\ (for $\Gamma=2$). We fit the three spectra simultaneously with spin tied between datasets (thereby implying a fixed inner edge, assumed to be at the ISCO as expected for a source in the soft state). The results of the spectral fitting and parameters
are given in Table 1, with a best-fitting value for the spin,
$a_*$~=~0, constrained to be $<$~0.004 (90\% error). We then changed
the inclination to 39\degree\ ($\Gamma=5$) and re-fitted the
data, finding the spin to reduce to $a_* < 0.002$.  However, in both cases the fit qualities are statistically very poor, implying that one or more of our assumptions about the model parameters is incorrect. This is somewhat unsurprising in light of the caveats involved with estimating the inclination via Doppler boosting (e.g. our assumption that the oscillation events should have a standard brightness temperature).

\subsection{Free inclination fits}
In the preceding analysis, we used the inclination angle derived from our
comparison of the short-timescale flaring to the analogous oscillation
events in GRS~1915+105. This implicitly assumed that the short
timescale radio variability seen in the M31 source was intrinsic; if
this was instead due to interstellar scintillation (M+13) then it cannot be
used to constrain the inclination angle. To account for this
possibility, we repeated the model fitting allowing both the spin and
inclination to be free, finding the best fit shown in Figure 1 (along
with the spin/inclination $\chi^2$ contours) and parameters given in Table
1. The now considerably better (and statistically acceptable)
best fit implies a spin of 0 at an inclination angle of 15\degree, and a 3$\sigma$ upper limit to the spin of 0.15, which occurs for an inclination angle of 0\degree. Although we may expect radio-selected sources to be at low
inclinations (and therefore brightest), it is worth noting that the
source was in fact X-ray selected (our observations were triggered by
the X-ray outburst reported by Henze et al. 2012) and if we assume a uniform distribution of
inclinations for the population of microquasars in M31, such a low
inclination is highly improbable. 

  \begin{figure*}
\centering
  \includegraphics[width=14cm]{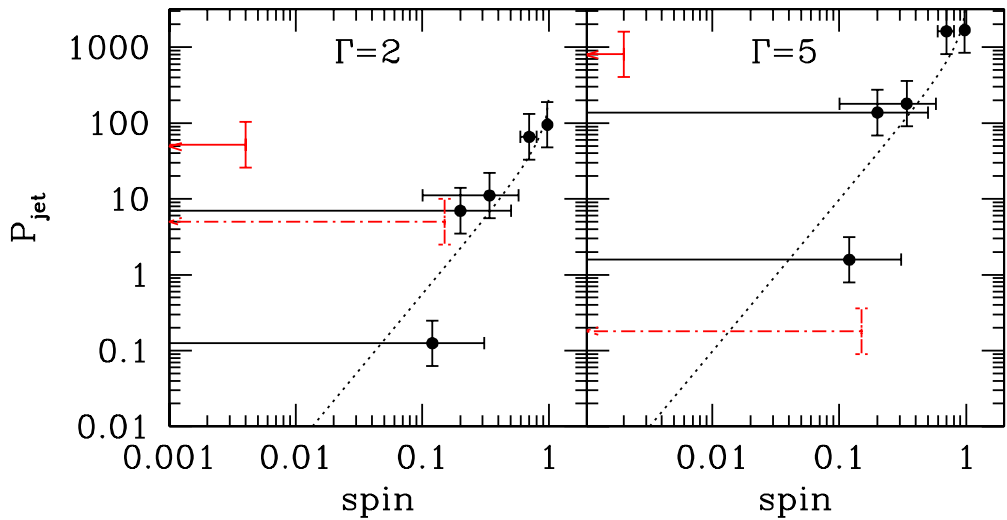}
\caption{{\it Left:} $P_{jet}$ (with units of GHz Jy kpc$^2$ M$_{\odot}^{-1}$) versus prograde spin, plotted
  for 5 BHBs (S13) and the microquasar in M31 (red lines). For the latter we have
  assumed a boosting factor (i.e. $\delta^{3}$) of $\approx$15 (M+13), $\Gamma$ = 2 and we have obtained $P_{\rm jet}$ values for $\alpha = -0.4$, as appropriate for ballistic
  ejections (NM12). The best fit to the five Galactic BHBs is
  shown (dotted line; $P_{\rm jet} \propto (M\Omega_{\rm H})^{2}$). As
  there is some uncertainty over the inclination we also plot the
  $3\sigma$ upper limit from the model fits with both spin and
  inclination left free (dot-dashed line). To be consistent with the best-fitting BZ relation, the inclination is required to be improbably low ($\approx$ 0\degree). {\it Right:} Colours and lines as for the left-hand plot but with values derived using $\Gamma$ = 5. We note that, whilst the best-fitting spin value is still 0, the points with free inclination are consistent with the best-fitting relation of S13 within 3$\sigma$ limits.}
\label{fig:l}
\end{figure*}

Although we are confident that our identification of Eddington behaviour and therefore mass estimate is reliable, we also investigate the effect of a representative 10\% error in the flux estimate at the peak of the outburst which would lead to a 10\% error in the mass, i.e. an upper limit of 11$M_{\odot}$. We once again find that in all cases the best-fitting spin is 0 with the largest 3$\sigma$ upper limit being 0.23, obtained when fitting with free inclination (with a best fitting value of 24$^{+4}_{-14}$\degree).

\subsection{A retrograde spin?}
In all cases the best fitting value of the spin is zero and constrained to be extremely low. As the value tends to the limits of the model this is a likely sign that the model we have used is simply incorrect in this case. Indeed, we had not considered whether the spin could be retrograde (a$_{*} < 0$), as such values have not been reliably reported (although we note the recent paper by Reis et al. (2013), whose iron line fitting leads them to claim a truncated disc or retrograde spin in a Galactic BHB - see also Kolehmainen, Done \& Diaz-Trigo 2013 for possible caveats). We used an extended version of the {\sc bhspec} model\footnotemark\footnotetext{bhspec\_spin\_0.01.fits, available from \\http://www.cita.utoronto.ca/$\sim$swd/xspec.html} that expands the spin parameter space down to -0.998, and re-fitted the data as before for the fixed inclinations and mass of 10$M_{\odot}$. The resulting fits are a highly significant improvement over the prograde fits and are given in Table 2. These remarkably imply well-constrained retrograde spin in all cases. Given the aforementioned caveats on deriving the inclination, we also fitted the data with both inclination and spin free to vary. The resulting best fit is shown in Figure 2 (with parameters and 90\% errors given in Table 2) and implies maximal retrograde spin, constrained to be $<$-0.57 (90\%) although we caution that once again the best fit tends to the limit of the model.

Retrograde spin appears to be statistically favoured in all cases (up to $\Delta\chi^{2}$  = 342 for a fixed inclination of 39\degree). However, from studying the residuals in Figure 3, we can see that the subtle changes in excess $\chi^{2}$ between the prograde and retrograde fits with free inclination occur in an energy range where the ratio between the models corresponds to only a $\sim$5\% difference, of the order of the systematic uncertainties in the model (Figure 3). So whilst retrograde spin is statistically favoured, we cannot rule out the spin being extremely low (or zero) when the inclination is left free.

\begin{figure*}
\centering
  \includegraphics[width=10cm]{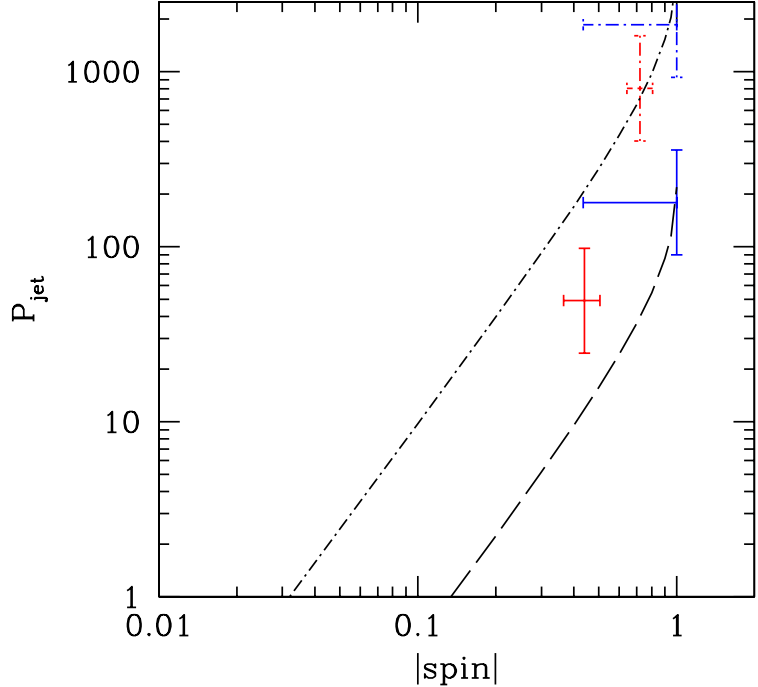}
\caption{$P_{jet}$ (with units of GHz Jy kpc$^2$ M$_{\odot}^{-1}$) versus $|$spin$|$, where we use the results of the spectral fitting which infer a retrograde spin (Table 2). The solid error bars and dashed lines (the latter being the best-fitting relations derived for prograde spin) are $\Gamma$ = 2 and dot-dashed are $\Gamma$ = 5. Red lines are jet powers inferred assuming a scaling of brightness temperature in the flares to those of GRS~1915+105. We also show the best-fitting spin value and inferred jet power where the inclination and spin are free to vary (blue lines). We note that, in the event that the scaling (and inferred boosting) is inaccurate, this is the most reliable measure. In all cases the points appear broadly consistent with the BZ effect producing similar jet powers in the retrograde regime as the prograde regime, without the need to invoke larger amounts of magnetic flux (Garofalo 2009).}
\label{fig:l}
\end{figure*}

\section{Blandford-Znajek powering of jets}

Given the extreme low spins resulting from the model fitting, the derived
values for the peak jet power are extremely useful, since they can
constrain whether the BZ effect is the dominant mechanism for
launching such ballistic jets.

Using the approach of NM12 we can estimate the maximum
jet power from the brightest radio flare of the few-month long flaring
sequence that was seen {\bf after} the initial rise to Eddington and
transition to the disc-dominated state. In the M31 microquasar, this
brightest flare was detected $\gtrsim60$\,days after the state
transition by AMI-LA at 15~GHz, at a flux density of $0.96\pm0.09$~mJy
(M+13). Such a bright radio flare could only have been produced by the
ballistic jets, or via interactions with the surrounding medium.
Although the latter can result in external shock acceleration and
therefore bright radio emission downstream in the jets, such flares
are seen to occur several months to years after the original outburst
and tend to be long-lived (e.g. Corbel et al. 2002), whereas the
flares we observed decayed on timescales of days.  Furthermore, the
lack of resolved emission in the VLBA image taken only 3 days later
rules out bright external shocks further than $\sim10^{16}$\,cm
downstream, significantly closer to the source than external shocks
have previously been observed to form in Galactic sources
(e.g.\ Corbel et al. 2002, 2005).

%The samples of NM12 and S13 explicitly include only those sources that
%reach $\ge$50\% Eddington in order to ensure the measured jet power
%corresponds to that of the ballistic jets (which have been suggested
%to be `standard candles'). 

In order to make a direct comparison to NM12, we scale the flux of the
AMI-LA flare to 5~GHz assuming a spectral index $\alpha$ (where S$_{\nu}\propto\nu^{\alpha}$) and use the representative value of $-0.4$ (based on the analysis of
other BHBs undergoing these events: NM12), giving a
scaled 5~GHz flux of 1.49 mJy. The jet power is then determined using
the proxy outlined in NM12: $P_{\rm jet} = \nu S_{\nu} d^{2} M^{-1}$,
where $S_{\nu}$ (Jy) is the beaming-corrected flux density,  $d$ is the distance in kpc and $M$ is
the mass in units of $M_{\odot}$. From the aforementioned values for
the mass and inclination, we
obtain the jet power values (with units of GHz Jy kpc$^2$ M$_{\odot}^{-1}$) given in Table 3. To compare to the 5 BHBs presented in S13 we also re-derive their de-boosted jet powers using the flux densities, inclinations, distances, masses and $\alpha$ values given in NM12 and S13; these jet powers are also given in Table 3.  

 As $P_{\rm
  jet}\propto a_{*}^{2}$ is only valid in the low spin regime, we
follow S13 and derive the best fitting, BZ relation for the 5 Galactic BHBs using $P_{\rm jet}
\propto ({\rm M}\Omega_{\rm H})^2$ (Tchekhovskoy, Narayan \& McKinney
2010) where $\Omega_{\rm H}$ is the angular frequency of the event
horizon in natural units, $\Omega_{\rm H} = a_{*}/(2 M (1+\sqrt{1-a_{*}^2}))$. Following NM12 we assume an error of 0.3 dex (i.e. a factor 2) on
$P_{\rm jet}$ due to uncertainties in the jet/ISM interaction and
$\alpha$, and assume symmetric errors on $\Omega_{\rm H}$. We determine the coefficients of the linear fits (using least squares fitting) to the logarithmic points (i.e. log$P_{jet} = \log({\rm M}\Omega_{\rm H})^2 + \log A$, thereby avoiding the fit being dominated by the point with the lowest power and therefore lowest error). We find best-fitting values for the correlation ($\log A$) of 2.94$\pm$0.22 (90\%) and 4.19$\pm$0.22 (90\%) for $\Gamma$=2 and $\Gamma$=5 respectively. 

We plot the prograde spin and associated jet power values for all 6 sources, and for $\Gamma$~=~2 and 5, in Figure 4 with the best-fitting relations provided above. The plot shows that all points derived using the fixed inclinations calculated from Doppler boosting of the variability brightness temperature sit well above the relation. 
When the inclination is left free to vary, the points are consistent with the best-fit BZ relation to the Galactic BHBs, however, for $\Gamma$=2, 
an improbably small inclination is required (0\degree\ at 3$\sigma$). For the source to sit on the BZ correlation, the much faster, $\Gamma$=5 jet at more moderate inclinations of $<$20\degree (at 3$\sigma$ from inspection of Figure 1) would be required.

In Figure 5 we plot the best-fitting retrograde spins and corresponding jet powers with the extrapolated relation for prograde spin. Unlike for prograde spin, the points appear broadly consistent with BZ powering and do not appear to require larger amounts of magnetic flux (Garofalo 2009).

% for alpha = -0.4, and delta^3.4 = 14, we get i=27.8 for gamma=2  
% for alpha = -0.4 delta^3.4=70 we get i=10 and a<0.23 for gamma=2. 
% for alpha = 0 and delta^3 = 70 we get i=0 and a = 0< 0.14 <0.29
% for alpha = 0 and delta^3 = 14 we get i=24

%By assuming that the Eddington oscillation events have a common brightness we have explicitly assumed that spin does not have an affect. 

%However, retrograde spin could potentially increase the jet brightness by an order of magnitude (Garofalo 2009). In which case making comparisons to the events in GRS1915+105 (which has a well established prograde spin: Middleton et al. 2006; McClintock et al. 2006; NUSTAR) may be flawed. As such we adopted the most robust approach and fitted the data with both inclination and spin free to vary. The resulting fit (Table 2) implies maximal retrograde spin, constrained to be $<$0.57 (90\%). We plot the best-fit spin values with the corresponding jet powers in Figure 4 along with the BZ best-fitting relations of S13. 

\section{discussion}

The proximity of the microquasar in M31 is such that the observed data quality
is high during outburst (even during the decay phase), whilst the fractional error on the distance ($\sim$6\%) is considerably
smaller than for most Galactic sources (usually a factor $\sim$2;
Jonker \& Nelemans 2004). Using the best available constraints on the
inclination and mass, we can confidently fit the three {\it
  XMM-Newton} spectra obtained in the decay phase using an
accretion disc model that incorporates the effect of
spin (Davis et al. 2005). For the mass inferred for the source,
we find the spin to be extremely low at $a_{*}<$~0.01 (90\%) if we assume only prograde spin to be allowed. This is much
lower than the majority of BHBs and is the lowest yet of
the `standard candle', Eddington BHBs (S13). However, given the poor quality of the spectral fits, it is clear that at least one of our assumptions is incorrect. By allowing the inclination to be a free parameter in the model we obtain a considerably better fit and still find the spin to be very low: a$_{*} <$ 0.06 (90\%). 

Taking the brightest flare after the source reached outburst
maximum (to be consistent with the approaches of NM12 and S13), seen
by AMI-LA at 15~GHz, we infer that the source sits above the best fit
BZ correlation - inferred from $P_{jet}$ and ${\rm M}\Omega_{\rm H}^2$ for the Galactic BHBs (S13) - for all points resulting from fixed inclinations in the spectral fitting (Figure 4). For free inclinations, the resulting jet power and spin are consistent with the relation for $\Gamma=2$ but only at the 3$\sigma$ limit of the permitted values and only for an improbably small inclination. At $\Gamma=5$, the values are consistent with the 
relation for more moderate inclinations ($<$20\degree) but whether such fast jets could be launched through BZ powering remains uncertain (see e.g. Barkov \& Komissarov 2008 who find $\Gamma$ rarely exceeds 3 for simulations with a$_*$=0.9). The lack of radio coverage during the rise to X-ray outburst prevented observation of a potentially brighter radio flare associated with the state transition. As a result we caution that the inferred jet powers may be lower limits which would enhance the deviation from the reported correlation. Taken as a whole, the evidence implies that
the BZ effect may not be the dominant mechanism for powering these
events for such low spin BHs, consistent with the analysis of Russell et al. (2013) and AGN studies of Sikora et al. (2013) and Sikora \& Begelman (2013). Alternatively the topology of the
magnetic field in the vicinity of the black hole may differ in this
source, resulting in a different scaling of $P_{\rm jet}$ with $a_*$
(Beckwith 2008; McKinney \& Blandford 2009).

If we allow the spin to be retrograde then we find that the spectral data statistically favour this scenario, although when the inclination is a free parameter, the 
difference between prograde and retrograde best-fit models (a$_{*}$ = 0 versus -0.998) is of the order of the model systematic uncertainties. Should the spin be truly retrograde then we find that the jet power is broadly consistent with the best-fitting relations for prograde spin, making it plausible (although not conclusively demonstrating) that the BZ effect may be producing these events. This does not appear to require an order of magnitude increase in jet power (Garofalo 2009) and could feasibly be supporting a scenario where the disc is arrested by magnetic fields (Tchekhovskoy \& McKinney 2012 and references therein). Once again, should there have been a larger, unobserved, flare at the hard-to-soft state transition then this may not be a valid supposition and, whilst a compelling result, the model systematics currently prevent a fully robust identification of retrograde spin. 

Although it is uncertain how such retrograde spins could occur in close binary systems, it has been suggested that wind-fed accretion can produce counter-aligned inflows (e.g. GX~1+4: Chakrabarty et al. 1997 and Cygnus X-1: Shapiro \& Lighman 1976, Zhang et al. 1997). We effectively rule this out here as the very low optical luminosity limit (M+13) precludes a giant companion star. It is plausible that a highly anisotropic supernova kick could displace the black hole relative to the secondary although it is not clear how this would then stabilise and lead to the required system geometry. A third possibility, presented by Reis et al. (2013) to explain the potentially retrograde spin of the black hole in Swift J1910.2-0546, is the tidal capture of a star (Fabian et al. 1975) after the black hole has formed whilst in a Globular Cluster (GC), with the natal kick velocity carrying the system out of the GC to its present location. The initial capture can produce a retrograde orbit and, as the timescales for alignment of the secondary are very long (e.g. Maccarone 2002) such systems will remain retrograde for much of their lifetimes. The nearest bright GC to our source is approximately 1.4~kpc away in projected distance (Bol D091, e.g. Rey et al. 2009). Given standard natal supernova kicks (e.g. Fabian et al. 1975; Mirabel et al. 2001; Gonz\'{a}lez Hern\'{a}ndez et al. 2006; see also Miller-Jones et al. PASA submitted) of a few tens to hundreds of km~s$^{-1}$ this would imply only a few million years of travel post-capture, which is entirely consistent with typical LMXB lifetimes of a few Gyr.  While this scenario is plausible (although of low probability), we cannot establish its likelihood in the absence of any constraints on the space velocity of the system.

\section{Conclusion}

We find ourselves with three possibilities to explain the bright jet events seen from this source. The first is that the spin is prograde and the BZ effect does not seem to power the events unless the inclination is (improbably) low, the jet is remarkably fast for such a low spin, or the magnetic field topology in this source is unusual. The second possibility is that, whilst of low probability, the spin is significantly retrograde and that the BZ effect may be responsible, with the suggestion that, if we truly observed the brightest radio flare then a magnetically arrested disc may be present. The third possibility is that our assumed mass is incorrect and the Eddington phenomenology can occur at lower mass accretion rates (i.e. for a higher BH mass). Whilst we cannot rule out this last possibility it would seem remarkable that the same, distinct behaviour is seen in both this source at sub-Eddington rates and in the most extreme Galactic BHB, GRS~1915+105, widely accepted to accrete at its Eddington limit when brightest. The latter issue will be resolved should a dynamical mass measurement be made for the black hole, yet this will require the next generation of extremely sensitive 40m class telescopes, i.e. ELT.

%In addition, we
%have also assumed a Lorentz factor of 2, however, even if the jet is
%travelling with $\Gamma$=5 (0.98c), the inclination only changes by
%small amount (21-22 degrees for $\alpha$=0 and -0.4 respectively) and
%the spin remains $<$0.1.

%i=20.65 alpha=0, gamma=5 d^3 = 14
%i=22.1 alpha=-0.4, gamm=5, d^3.4=14

Given the uncertainties inherent in comparing such extreme
sources, it is imperative that we both study the varied flaring
properties of known (and new) Galactic sources and discover more microquasars by
searching in nearby galaxies (Middleton et al. in prep). Once
discovered, such sources will allow increasingly robust tests for the
BZ effect, and, where multi-wavelength campaigns are carried out and
conditions are favourable (e.g. source inclination), it
may be possible to probe the disc-jet coupling directly and determine
the underlying physics of the launching of relativistic jets.

\section{Acknowledgements}
The authors thank the anonymous referee for helpful suggestions, and Chris Done, Shane Davis and Jason Dexter for useful discussion. MJM acknowledges
support via a Marie Curie FP7 Postdoctoral scholarship and JCAMJ
acknowledges support from an Australian Research Council Discovery
Grant (DP120102393). This work is based on observations obtained with
{\it XMM-Newton}, an ESA science mission with instruments and contributions
directly funded by ESA Member States and NASA. We also thank the staff
of the Mullard Radio Astronomy Observatory for their assistance in the
commissioning and operation of AMI, which is supported by Cambridge
University and the STFC.

\label{lastpage}

\end{document}